\documentclass{ptptex}

\usepackage{wrapft}
\usepackage{graphicx}
\usepackage{amsmath}
\usepackage{amssymb}
\usepackage{amsfonts}
\usepackage{amsthm}
\usepackage{amsxtra}
\usepackage{mathrsfs}

\upshape

\def\bra#1{\langle #1 |}
\def\ket#1{| #1 \rangle}
\def\Mod#1#2{{{}_{#1}\ModM_{#2}}}

\def\1half{\frac{1}{2}}

\def\ap{{\alpha'}}

\def\Ub{{\bar{U}}}
\def\Vb{{\bar{V}}}

\def\Ab{{\bar{A}}}
\def\Cb{{\bar{C}}}

\def\zb{{\bar{z}}}

\def\crea{a^\dagger{}}
\def\anih{a}

\def\Sb{{\bar{S}}}
\def\Tb{{\bar{T}}}

\def\E{\mathrm{e}}

\def\Tr{\mathrm{Tr}}

\def\AlgA{\mathcal{A}}
\def\ModM{\mathcal{M}}
\def\OpO{\mathcal{O}}

\def\Cplx{\mathbb{C}}

\def\Hil{\mathscr{H}}
\def\Lag{\mathscr{L}}
\def\Intg{\mathbb{Z}}

\title{Solution Generating Technique for Noncommutative Orbifolds}

\author{Kento ICHIKAWA
  \footnote{ E-mail: kent@hep-th.phys.s.u-tokyo.ac.jp}}
\inst{  Department of Physics,
  University of Tokyo Tokyo 113-0033 ,Japan\\
  (Recieved March 11, 2002)}
\date{}
\abst{
 We propose the relationships between the noncommutative solitons and
 the (fractional) D-branes on the $\Cplx^2/\Intg_N$ orbifold and extend
 the solution generating technique for the orbifold. As applications, we
 determine how tachyon condensations occur in various D-\={D}
 systems on the orbifolds. The calculations give results consistent with
 BSFT. The extended solution generating technique enables us to
 calculate more general decay modes of D-\={D} systems.}

\begin{document}
\maketitle
\section{Introduction}

In recent years, nonperturbative methods of string theory have
revealed some mysterious dynamical structures through investigation of
the Sen's conjectures .\cite{Sen} conjectures are as follows. When the
tachyon of an unstable systems, such as a D-\={D} systems, is
condensed to the minimum of the tachyon potential, the brane is
reduced to a lower dimensional brane or is annihilated. Whether a
lower dimensional brane remains after tachyon condensation and how
lower dimensional brane remains is depend on the topological charges
of the gauge field on the D-brane before the condensations.

Recently noncommutative geometry has attracted
grate interest as a low energy effective theory of open string
theory and the theory of D-branes. The philosophy of noncommutative
geometry is that an algebra of the functions on a manifold rather than a
set of points exist {\it a priori}.

Grate progress was recently made in noncommutative field theory \cite{SW}
(for review, see {\it e.g.} Ref \cite{DN1} by the discovery of GMS soliton
\cite{GMS}. It is interpreted as D-brane, \cite{HKLM} a in accordance with the
calculation of tension and the observation that the fluctuation modes
around the solitonic solution are identical with the spectrum of the
D-brane. Owing to this development, noncommutative field theory has
become one of the most powerful tool to extract information about
nonperturbative aspects of string theory.

Noncommutative field theory possesses powerful techniques for analyzing
nonperturbative aspects of string theory, such as the solution generating
technique \cite{HKL} -- which enables us to generate another solution
from a (in most case trivial) solution -- and the methods of reading
off the topological charge from the tachyon configuration
cite{Mat,HM,Wit1}. In particular, noncommutative version of the field
theory obtained using BSFT \cite{KL,TTU} enables us to analyze tachyon
condensations in D-\={D} systems in the language of noncommutative
geometry. There are many other methods to analyze D-\={D} systems such
as BSFT \cite{Wit2} and cubic SFT \cite{Wit3}-like approach
\cite{SSFT,Ber} (for review, see Ref \cite{Oh}), and some interesting results
have gotten \cite{Sen2,MS,NTU,Hori}.  Furthermore because noncommutative
geometry can be defined on various types of manifolds, these methods can
be applied to various backgrounds such as tori \cite{KMT,KS}, fuzzy
spheres \cite{HNT} or orbifolds \cite{MM}.

In this paper we concentrate on $\Cplx^2/\Intg_N$ orbifolds. These
orbifolds have some peculiar features \cite{DM} for example, \cite{MM} that they
can have fractional D-branes whose R-R charges are fractional.  In Ref
\cite{MM}, a general framework for noncommutative field theory on the
orbifold was given and some simple calculations were made. Here we
develop this framework for application to nontrivial configurations
which Ref \cite{MM} did not deal with using some additional proposals of
identifications between noncommutative solitons and (fractional)
D-branes and extending the solution generating technique. This enables
us to calculate more general classes of D-\={D} decays. As we see
below, we can construct $N$ types of noncommutative solitons, and we
propose some rules to identify these solitons as fractional D-branes
of different types. Using this identification and by extending the
solution generating technique, we can calculate various new classes of
decay modes of D-\={D} systems.  Calculations of the decay of D-\=D
systems using BSFT are given in Ref \cite{Tak1}. We carry out explicit
calculations using noncommutative field theoretical methods in some
simple cases. We also show that in some tachyon configurations, D-\=D
systems decay with extra D-\=D pairs and the extra D-\=D pairs are
annihilated. We calculate general two pair of D-\=D systems, and show
that this system decays into D0-branes and D2-branes. Our method is
applicable to analysis of more general systems.

We review noncommutative field theory on orbifolds $\Cplx^2/\Intg_N$ in
\S\ref{NCFT}. In \S\ref{condensation}, we calculate decay
modes of various D-\={D} systems in $\Cplx^2/\Intg_N$ whose codimensions
are two and four. Finally we give some conclusions and discussions in
\S\ref{conclusion}.

\section{Noncommutative field theory on noncommutative orbifold}
\label{NCFT}

\subsection{The algebra of noncommutative orbifold}
\label{OrbAlg}

To formulate field theory on $\Cplx^2/\Intg_N$ orbifolds, we must know
about the algebra of functions on noncommutative orbifolds. To start
with, we consider the noncommutative algebra of $\Cplx^2$. We assume
that the noncommutativity of the coordinates satisfy
\begin{align}
 \begin{split}
  [z_1,\zb_1]=\theta ,\quad &\,[z_2,\zb_2]=\theta, \\
  [\anih_1,\crea_1]=1,\quad &\,[\anih_2,\crea_2]=1, \\
  \mbox{where}\anih_1=z_1/\sqrt{\theta},\quad&\,\anih_2=z_2/\sqrt{\theta}.
 \end{split}
\end{align}
We set the commutator of the $z_1$ plane and $z_2$ plane to the same
value for the ease of calculation.  The algebra of noncommutative
$\Cplx^2$ is generated by the operators [ or tensor products of two
matrices ] acting on the tensor product of two Fock spaces. We 
denote its basis as $\ket{n_1}\otimes\ket{n_2}$ and abbreviate it as
$\ket{n_1,n_2}$ where $n_1 $and $n_2$ are non-negative integers.

The generator of the orbifold group acts on the coordinates as
\begin{equation}
 z_1\rightarrow \E^{\frac{2\pi i}{N}} z_1,\quad
  z_2\rightarrow \E^{-\frac{2\pi i}{N}} z_2.\label{orb-action}
\end{equation}
Following the prescription given in Ref \cite{MM}, the algebras which
represent the orbifolded space is ``covariant'' ({\it invariant}) in the
sense of Refs \cite{MM} and \cite{DM}) under the action of the orbifold group
(\ref{orb-action}). The algebra decomposes into $N$ subalgebras, and these
subalgebras are related by the automorphisms defined by
(\ref{orb-action}). This situation reminds us of the fact that in boundary state
construction of the D-brane on the orbifolds that there are $N$ boundary
states on the noncommutative orbifolds related to each other.

With the decomposition of the algebra, the Hilbert space (Fock space
in our case) are also decomposed. The components of the decomposed
Hilbert spaces transform under different representations of the
orbifold group $\Intg_N$:
\begin{align}
 \begin{split}
  \Hil &= \bigoplus_{r=0}^{N-1} \Hil_\alpha \\ \Hil_\alpha
  &=\ket{n_1,n_2}\quad (n_2-n_1=\alpha\,\,\mbox{mod $N$}).
 \end{split}
\end{align}
The subscript $\alpha\,(\alpha\in\Intg/\Intg_N)$ indicate that corresponding
conponent of the decomposed Hilbert space transform under the $\alpha$-th
representation. That is, if
$v\in\Hil_\alpha$ then $v$ transforms as $v\rightarrow\E^{2\pi
i\alpha/N}v$. The decomposition of the Hilbert space clearly reflects the
decomposition of the algebra. Furthermore, we see that there are
parts of the algebra that are not ``covariant''. These parts of the
original algebra are bimodules which connect the decomposed Hilbert
spaces. We refer to these bimodules $\Hil_\alpha$ and $\Hil_\beta$ as
$\Mod{\alpha}{\beta}$.

Bimodules $\Mod{\alpha}{\alpha}=:\AlgA_\alpha$ are ``covariant'' algebras describing
a noncommutative orbifold and all algebras of this form are identical.
$\Mod{\alpha}{\beta}\, (\alpha\neq\beta)$ are not algebras but bimodules
on which the orbifold algebra $\AlgA_\alpha$ acts from the right and
$\AlgA_\beta$ acts from the left.

When we specify the algebras which act from the left and right, we can
redefine $\Mod{\alpha}{\beta}$ as
$\bigoplus_{\gamma=0}^{N-1}\Mod{\alpha+\gamma}{\beta+\gamma}$. This
redefinition of the bimodules implies a change of the orbifold
algebra. The new orbifold algebra is direct sum of old orbifold
algebras: $\bigoplus_{\gamma=0}^{N-1}\AlgA_{\alpha+\gamma}$. 
%One may
%suspect that this redefinition leads us to wrong results because we
%change the algebla by hand. But as far as we specify the Hilbert space
%which the algebra acts on, the direct sum component which works is
%restricted to only one component and causes no problem.  This
%redefinition is allowed only when we specify the Hilbert space which the
%bimodule or algebra acts on, therefore index which specify the Hilbert
%space is indispensable.

Note that although direct sum components are introduced by hand, and
therefore cannot act on the specified decomposed Hilbert space, these
components have some meanings, when we interpret noncommutative
soliton as fractional D-brane below. We do not know why this
redefinition has meanings. Roughly speaking, we conjecture that this
is because string theory ``sees'' the covering space of the orbifold
by considering the twisted sector.

The explicit form of a redefined $\Mod{\alpha}{\beta}$ is the tensor
product of two matrices acting on the two Fock spaces whose coefficient
of the element
\begin{equation}
 \ket{\gamma+pN+q}\bra{q}\otimes\ket{\gamma+\beta-\alpha+rN+s}\bra{s},
\end{equation}
\begin{equation}
  \mbox{where} p,q,r,s,\gamma\in\Intg
   \quad\gamma+pN+q,\gamma+\beta-\alpha+rN+s,q,s\geq 0,
\end{equation}
can be nonzero, and all the other elements are zero. Note that $\Mod{
\alpha}{ \beta}\cong\Mod{ \alpha +\gamma}{\beta +\gamma}$ for any
$\gamma$.

The product of these operators is simply defined as the product of each
matrix:
\begin{equation}
 M_1\otimes M_2 \times N_1 \otimes N_2 := (M_1 N_1)\otimes (M_2 N_2).
\end{equation}
This structure of the product comes from the multiplication rule of
commutative functions.

\subsection{Noncommutative field theory and fractional D-branes}
\label{frac}

Field theory on noncommutative orbifolds is almost the same as
that on the noncommutative Euclidean space. However, there are some
differences, as discussed below:
\begin{enumerate}
 \item The normalization of the trace has a factor of $1/N$, because
       overall volume is $1/N$ while the identity operator in
       $\AlgA_\alpha$ is the same form\footnote{It is not clear whether we
       can say identity operator of $\AlgA_\alpha$ has the same form as
       that in noncommutative Euclidean space, because when we defined
       $\AlgA_\alpha$ we specified the direct sum component of the
       Hilbert space on which $\AlgA_\alpha$ acts and the Hilbert space
       is not the same one as in the case of Euclidian space. However
       we conjecture that this procedure is correct when we analyze string theory,
       because of the existence of the twisted sector.}. Thus the
       integration over the orbifolds is written
       $\frac{(2\pi\theta)^2}{N}\Tr$.
 \item Because the orbifold algebra is not an algebra of all infinite size
       matrices but, rather $\AlgA_\alpha$, the symmetry of the theory is not
       $U(\infty)\otimes U(\infty)$, but insted its subgroup $G\subset
       U(\infty)\otimes U(\infty)$ which preserves $\AlgA_{\alpha}$:
       \begin{equation}
	g\AlgA_{\alpha} g^\dagger = \AlgA_{\alpha}.\quad(g\in G)
       \end{equation}
\end{enumerate}

Considering the correspondence between noncommutative solitons and
D-branes in the Euclidean space, it is natural to identify
noncommutative solitons on an orbifold as (fractional) D-branes. In this
subsection we will give the rules of identification and give some
evidences for this identification.

In the following several paragraphs, we explain how to determine the
type of noncommutative solitons and D-branes. Noncommutative soliton
on the D4-brane wrapping on the $\Cplx^2/\Intg_N$ orbifold
\footnote{When we refer to the D4-brane or the D2-brane below, we are
always referring to branes wrapping the orbifold and having twisted
R-R 2 or 4 form charges, {\it not} transverse to the orbifold like in
\cite{DM}} is characterized by the projection operators
$P\in\AlgA_\alpha$, which satisfy $PP=P$. When we consider the tachyon
condensation of the D-\=D system, the projection operator is $1-T\Tb$
or $1-\Tb T$, where $T$ is the tachyon field.

If a noncommutative soliton has the form $\ket{k}\bra{k} \otimes
\ket{l}\bra{l}$, we call $l-k$ (modulo $N$) the type of noncommutative
soliton. The reason for this classification is as follows. When $n$
noncommutative solitons of different types coincide, the gauge
symmetry is not enhanced to $U(n)$ but remains $U(1)^n$. This is
because the open strings connecting solitons of different types are
expressed by nontrivial representations of the orbifold group and are
prohibited as a gauge field. Thus these noncommutative solitons of
different types should be distinguished.

\begin{wrapfigure}{l}{6cm}
%\begin{figure}
\begin{center}
 \begin{picture}(0,0)%
\includegraphics{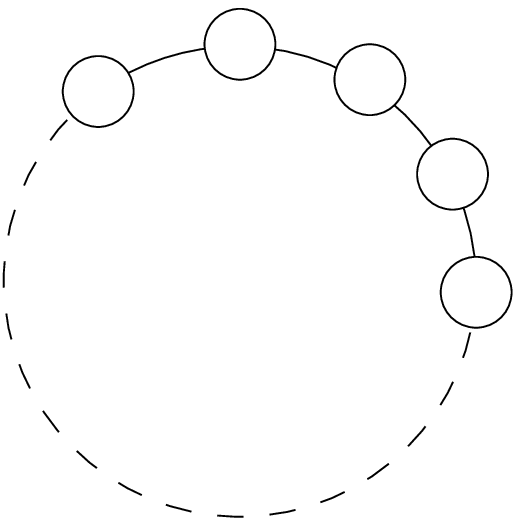}%
\end{picture}%
\setlength{\unitlength}{4144sp}%
\begingroup\makeatletter\ifx\SetFigFont\undefined%
\gdef\SetFigFont#1#2#3#4#5{%
  \reset@font\fontsize{#1}{#2pt}%
  \fontfamily{#3}\fontseries{#4}\fontshape{#5}%
  \selectfont}%
\fi\endgroup%
\begin{picture}(2339,2338)(406,-1636)
\put(793, -7){\makebox(0,0)[lb]{\smash{\SetFigFont{10}{12.0}{\familydefault}{\mddefault}{\updefault}
$N-1$
}}}
\put(1387,209){\makebox(0,0)[lb]{\smash{\SetFigFont{10}{12.0}{\familydefault}{\mddefault}{\updefault}
$0$
}}}
\put(1873, 47){\makebox(0,0)[lb]{\smash{\SetFigFont{10}{12.0}{\familydefault}{\mddefault}{\updefault}
$1$
}}}
\put(2143,-331){\makebox(0,0)[lb]{\smash{\SetFigFont{10}{12.0}{\familydefault}{\mddefault}{\updefault}
$2$
}}}
\end{picture}
\end{center}
\caption{The quiver diagram for gauge fields on D-branes wrapping on
 $\Cplx^2/\Intg_N$. Each node represents a fractional D-brane. The numbers
 represent the types of D-branes associated with the irreducible
 representation of the orbifold group $\Intg_N$. We can rotate and
 overturn this quiver diagram before numbering the nodes.}
%\end{figure}
\end{wrapfigure}

Next, we define the types of D-branes. We refer to the positions of the
nodes in the quiver diagrams for gauge fields on $\Cplx^2/\Intg_N$ as the
``types'' of the D-branes \cite{DM}. The positions of nodes in the
quiver diagrams are related to the irreducible representations of the
orbifold group $\Intg_N$. Thus we can regard the types of the D-branes
as irreducible representations of the orbifold group \cite{JM,BCR,DG}.
We note that the important quantity for types of (fractional) D-branes
is the absolute value of the difference between the types of two
(fractional) D-branes, because we can rotate and overturn the
diagram. In boundary state language \cite{BCR}, this fact is more
obvious, because the difference of the two D-branes is just a phase
factor $\exp( 2\pi i \alpha / N)$, in the string theory on
$\Cplx^2/\Intg_N$. To fix these degrees of freedom, we identify the
types of the D-branes by the integer $\alpha [=0,\cdots,N-1]$, where
the generator of the irreducible representation is $\exp (2\pi i
\alpha /N)$. The elements of $\AlgA_\alpha$ are functions on a
D4-brane \cite{MM}. We use the convention in which the type of the
D4-brane whose algebra $\AlgA_{\alpha}$ is $\alpha$.

We next propose rules of identification between noncommutative solitons on
the type $\alpha$ D4-branes and D0-branes or D2-branes as follows.
\begin{itemize}
 \item The type of a noncommutative soliton on the D4-brane of type
 $\alpha$ is determined by the following rule.\footnote{We can adopt
 another rule in which $\ket{k}\bra{k} \otimes I$ corresponds to $\alpha
 + \beta $ and $ I \otimes \ket{k}\bra{k}$ corresponds to $\alpha -
 \beta -1$. This is just a difference of convention, and the physics is
 the same.}\\
       \begin{enumerate}
	\item If the soliton has the form $\ket{k}\bra{k}\otimes I$ ($I
	\otimes \ket{k}\bra{k}$), where $k=\beta\,\, \mbox{(mod N)}$, the
	type of corresponding D2-brane is $\alpha+\beta+1$
	($\alpha-\beta$). These D2-branes extend in the $z_2$ ($z_1$)
	plane.
	\item If the soliton has the form
	$\ket{k}\bra{k}\otimes\ket{l}\bra{l}$, where
	$k=\beta,l=\gamma\,\, \mbox{(mod$N$)}$, the type of D0-brane to
	which the soliton corresponds is $\alpha +\beta -\gamma +1$.
       \end{enumerate}
\end{itemize}
These rules can be verified as follows. We will concentrate on the
second rule, since the first rule can be understood in the same
way. Consider two solitons $\ket{k} \bra{k} \otimes \ket{l} \bra{l}$
and $\ket{k'} \bra{k'} \otimes \ket{l'} \bra{l'}$. An open string that
connects the former soliton to the later is written as
$\ket{k}\bra{k'}\otimes\ket{l}\bra{l'}$ and transforms under the
generator of the orbifold group $\Intg_N$ as $\exp[2\pi
i((k-k')-(l-l'))/N]$. Thus the difference between the types of the two
D0-branes is $(-k+l)-(-k'+l')$. Furthermore if we consider an open
string that connects a soliton on an $\alpha$ type D4-brane to a
soliton of the same type on a $\beta$ type D4-brane the difference of
the types of these D0-branes should be $\alpha - \beta$, because the
open string is an element of $\Mod{\alpha}{\beta}$. Taking these
points into account, the rules of correspondence between types of
branes and types of noncommutative solitons is determined as above, up
to the relative sign of the differences. The relative sign of the
differences is determined in such a way that the calculation of D-\=D
decay gives consistent result with the result obtained using BSFT
\cite{Tak1}.

It is known that the D4-branes and D2-branes have the same tensions as
normal D4-branes and D2-branes respectively, while the tensions of
fractional D0-branes are $1/N$ smaller than the bulk D0-branes. This is explained
in noncommutative language as follows. A D0 brane has fractional tension
because of the $1/N$ factor of the trace. Also, the energy
of D4 and D2-branes must be divided by the volume of a brane that is
$1/N$ times as large as that in the $\Cplx^2$ case. Therefore, the $1/N$
factor is canceled and the tensions of D2 and D4-branes are the same as
those of bulk D2 and D4-branes.

When all $N$ types of the fractional D0-branes are coincident, the
entire system has the same tension as the bulk D0-brane. Thus we can
regard it as a bulk D0-brane, and we can move it from the origin, in
contrast to the situation for individual fractional D0-branes which
are pinned at the origin. This is known for the commutative case
\cite{DDG}. This can also be explained in the language of
noncommutative field theory. A short explanation is presented at the
end of \S\ref{solgen}.

\subsection{The action of the D-brane
and D-\={D} systems}

Since we will consider the gauge theory on the orbifold
$\Cplx^2/\Intg_N$, we need operator representations of the covariant
derivatives. These are given as
\begin{align}
 \begin{split}
  C_{1,2}=\crea_{1,2}+i\theta^{1/2}A_{1,2},\quad
  \Cb_{1,2}=\anih_{1,2}-i\theta^{1/2}\Ab_{1,2}.
 \end{split}
\end{align}
$A_{1,2}$ and $\Ab_{1,2}$ are gauge fields on the orbifold. Because
$A_{1,2}$ and $\Ab_{1,2}$ must be invariant under the action of orbifold
group, they are elements of modules, $A_1 ,\Ab_2
\in \Mod{ \alpha -1}{ \alpha}$ and $A_2 ,\Ab_1 \in \Mod{ \alpha +1}{
\alpha}$. The reason is that the $\exp{(\mp 2\pi i/N)}$ factor resulting from the
action of orbifold group must be canceled by the contribution from the
rotation of the vector $A_\mu\rightarrow R(g)_\mu^\nu A_\nu$, where
$A_\mu [\mu=1,2,3,4]$ is the gauge field in the $x_\mu$ basis and
$R(g)_\mu^\nu$ is the matrix representing the rotation by the generator
of the orbifold group $\Intg_N$. Thus these gauge fields are invariant under
the action of the orbifold group.

The action of gauge theory on the noncommutative worldvolume of a
D4-brane that fills the orbifold is \cite{SW}
\begin{equation}
 \begin{split}
  T_{4}\left(\frac{2\pi \ap}{\theta}\right)^2&\int\frac{dx^4}{N}
  \left(-\frac{1}{4}F_{\mu\nu}^2\right)\\
  =T_{4}\left(\frac{2\pi \ap}{\theta}\right)^2&\frac{(2\pi\theta)^2}{N}\Tr\Big(
  -\frac{\theta^{-2}}{2}\big(([C_1,\Cb_1]+I)^2+([C_2,\Cb_2]+I)^2\\
  &-[C_1,C_2][\Cb_1,\Cb_2]-[C_1,\Cb_2][C_2,\Cb_1]\big)
  \Big),
 \end{split}
\end{equation}
where $T_p$ is the tension of the Dp-brane. 

We need the action of the field theory on the noncommutative orbifold
for D4-\=D4 systems. This action is given as
\begin{equation}
 \begin{split}
  \Lag&=T_{4}\left(\frac{2\pi\ap}{\theta}\right)^2
  \frac{(2\pi\theta)^2}{N}\Tr\Big(f(T,\Tb)
  \mbox{(kinetic terms of gauge fields)}\\
  &+ \frac{i}{4}\theta^{-1}g(T,\Tb)\sum_{i=1,2}
  \big((C^+_i\Tb-\Tb C^-_i)(\Cb^+_iT-T\Cb^-_i)
  + (C^-_iT-TC^+_i)(\Cb^+_i\Tb-\Tb\Cb^-_i) \big)\\
  &+ V(T\Tb -I)+V(\Tb T -I)
  +\mbox{(higher derivatives)}\Big),
 \end{split}\label{D4-D4bar}
\end{equation}
where $C^\pm_{1}$ and $C^\pm_{2}$ are covariant derivatives on the D-brane (
superscript $+$ ) and anti-D-brane ( superscript $-$ ). The tachyon
field $T$ is a bifundamental field, and $T\in\Mod{\alpha}{\beta}$ if
the D4-brane is of type $\alpha$ and the anti-D4-brane is of type $\beta$.
It couples to the gauge field on both the D-brane and anti-D-brane
via covariant derivatives. The gauge field on the D-brane acts from the
right, and the gauge field on the \={D}-brane acts from the left. For $\Tb$,
the actions of gauge fields are reversed. The potential $V(T\Tb -I) +V(
\Tb T -I)$ is minimal at $|T\Tb|=1$ and takes the value 1 at
$|T\Tb|=0$. This is identical to the action obtained in the BSFT description
\cite{TTU}.

The transformation law of tachyon fields under the generator $g$ of the
orbifold group is given by
\begin{align}
 \begin{split}
  T\in\Mod{\alpha}{\beta}&\rightarrow\E^{\frac{2\pi i}{N}(\beta-\alpha)}T.
  \label{transformation}
 \end{split}
\end{align}

\section{Tachyon condensation on noncommutative orbifolds}
\label{condensation}

\subsection{Solution generating technique in orbifold theories}
\label{solgen}

We now give a brief explanation of the solution generating technique.
To begin with we work on two dimensional Euclidean space. When there
exists gauge symmetry, the action possesses $U(\infty)$ symmetry,
$\OpO\rightarrow g\OpO \bar{g}\,(g\in U(\infty))$. This $U(\infty)$
symmetry enables us to employ a technique called ``solution generating
technique''. To use this technique, we must use a partial isometry $U$. A
partial isometry is an infinite-dimensional matrix that satisfies
\begin{equation}
 \Ub U=1\, , \, U\Ub\neq 1.
\end{equation}
Matrices that satisfy this relation do exist. For example, the
shift operator,
\begin{equation}
 S=\sum_{k=0}^{\infty}\ket{k+1}\bra{k},
\end{equation}
is partial isometry, and it satisfies the following relation:
\begin{align}
 \Sb S=I,S\Sb=I-P_1,
\end{align}
where $P_n$ is the projection operator $\sum^{n-1}_0\ket{k}\bra{k}$. Using
this isometry, we transform fields that are solution of the equation of
motion as
\begin{equation}
 \OpO\rightarrow U\OpO\Ub.
\end{equation}
Because
\begin{equation}
 \frac{\delta S}{\delta\OpO}\rightarrow 
  U\frac{\delta S}{\delta \OpO}\Ub,
\end{equation}
the transformed fields also satisfy the equation of motion. Consequently,
using the isometry we can construct another solution from a given solution.

Below we apply the solution generating technique to
$\Cplx^2/\Intg_N$ orbifolds. But, a straightforward application of the solution
generating technique is not useful for following reasons. Because the
symmetry of the action is not $U(\infty)\otimes U(\infty)$ but $G$, we
should use a partial isometry related to $G$, which is complicated.
Furthermore, when the types of the D-brane and anti-D-brane are
different, we cannot take a trivial configuration as the start point of
the solution generating.

To make more use of the solution generating technique, we can extend the
technique to use the subset of the isometries related to $U(\infty)
\otimes U(\infty)$ which satisfies the condition given below;
\begin{align}
 \ket{n_1,n_2}&\mapsto\ket{n_1',n_2'},\label{map-a}\\
 (n_2'-n_1')-(n_2-n_1)&=\gamma\quad(\mathrm{mod}N)\quad\forall{n_1,n_2}\label{map-b}.
\end{align}
%partial isometry is applicable to generate solutions when ``shifting''
%of the partial isometry can be defined. ``Shifting'' can be defined when
%all the vectors of the form $\ket{n_1,n_2}$ are mapped by the partial
%isometry to the vectors of form $\ket{n_1',n_2'}$ and all
%$({n'{}_2}-n'{}_1)-(n_2-n_1)$ coincide modulo $N$. ``Shifting'' means
%the value of $({n'{}_2} -n'{}_1) -(n_2 -n_1)$ modulo $N$. 
We call $\gamma$ in the above equation ``shifting''.  Note that if
$U$ is a partial isometry whose ``shifting'' is $\gamma$, then
\begin{equation}
 \Ub U =I\otimes I,\quad U\Ub =I\otimes I - I\otimes P_{\gamma+pN},
 \quad(p\in\Intg)
\end{equation} 
or
\begin{equation}
 \Ub U =I\otimes I,\quad U\Ub =I\otimes I - P_{-\gamma+pN}\otimes I.
\end{equation}

The reason for including the condition given in \ref{map-a}and
\ref{map-b} is as follows. Consider the solution generating technique
on the orbifold $\Cplx^2/\Intg_N$ which is constructed from the
partial isometry whose ``shifting'' is $\gamma$. It takes the operator
$\OpO\in\AlgA_\alpha$ to $U \OpO \Ub \in \AlgA_{ \alpha +\gamma}$. The
set $\AlgA_{\alpha +\gamma}$ is isomorphic to $\AlgA_\alpha$, and the
actions of the operators in $\AlgA_\alpha$ and those of the operators
in $\AlgA_{\alpha+\gamma}$ have the same form. Thus the same arguments
given in the case of the Euclidean spaces regarding the solution generating
technique is can be made here too. The nontrivial point concerns covariant
derivatives. By the isometry that satisfies the above condition, the
covariant derivative transforms $\Mod{\alpha \pm 1}{\alpha}$ to
$\Mod{\alpha +\gamma \pm 1}{\alpha +\gamma} \cong \Mod{\alpha \pm
1}{\alpha}$. Because covariant derivatives take arbitrary values in
$\Mod{\alpha +\gamma \pm 1}{\alpha +\gamma} \cong \Mod{\alpha \pm
1}{\alpha}$, they are transformed into covariant derivatives of the brane of type
$\alpha+\gamma$. Consequently, the solution obtained with the solution
generating technique is the solution of the $\AlgA_{ \alpha +\gamma}$ system.

We use the partial isometry composed  of the shift operators:
\begin{align}
 \begin{split}
  S_1&=S\otimes I=\sum_{k=0}^{\infty}\ket{k+1}\bra{k}\otimes I, \\
  S_2&=I\otimes S=I\otimes\sum_{k=0}^{\infty}\ket{k+1}\bra{k}.
p \end{split}
\end{align}
These shift operators satisfy the condition given above. The
``shifting'' of $S_1$ and $S_2$ are $-1$ and $1$, respectively.

When we use the solution generating technique in the D-\={D} system,
we can act with different partial isometries on the D-brane and
anti-D-brane to generate new solutions. That is, from a given
solution, the transformation
\begin{align}
 \begin{split}
  \OpO^+&\rightarrow U\OpO^+\Ub,\\
  \OpO^-&\rightarrow V\OpO^-\Vb,\\
  T&\rightarrow VT\Ub
 \end{split}
\end{align}
generates another solution. For example, consider a solution of the
D-\=D system with the tachyon operator $T$ in
$\Mod{\alpha}{\beta}$. Consider the case that we act shift operator
$S_2$ on the D-brane of type $\alpha$ but keep the anti-D-brane of
type $\beta$ unchanged. Then we get another gauge configuration of the
D-brane and $T\in\Mod{\alpha+1}{\beta}$. This is a solution of the
$(\alpha+1)$-$\bar{\beta}$ system, because the action of the
$\alpha$-$\bar{\beta}$ system and the $(\alpha+1)$-$\bar{\beta}$
system are the same and the logic of solution generating uses
only a superficial form of the action.

As the starting point of the solution generating technique, we consider the
most trivial gauge configuration of the $\alpha$-$\bar{\alpha}$ system,
\begin{align}
 \begin{split}
  C_1^+&=\crea_1,\Cb_1^+=\anih_1, \\
  C_2^+&=\crea_2,\Cb_2^+=\anih_2, \\
  C_1^-&=\crea_1,\Cb_1^-=\anih_1, \\
  C_2^-&=\crea_2,\Cb_2^-=\anih_2, \\
  T&=I\otimes I.\label{start}
 \end{split}
\end{align}
This configuration is obviously a solution of the action of the D-\=D
system (\ref{D4-D4bar}), which represents the vacuum. We can generate
many solutions from this solution using the extended solution
generating technique.

This is a good place to give an explanation of how an
individual noncommutative soliton is pinned at the origin,while a collection of
$N$ branes of $N$ different types can move from the origin. The solution
generating technique by shift operators can be generalized as
\begin{equation}
 C_i\rightarrow S^n C_i\Sb^n + X_i,
\end{equation}
where each $X_i$ is an  $n\times n$ matrix \cite{HT}. The new degrees
of freedoms represented by these $X_i$s must be in the same representation of
the orbifold group as $A_i$. The action for $X_\mu\, (\mu = 1 - 4)$ is
$([X_\mu,X_\nu])^2$, and this action is minimized when all $X_i$
commute. It is known that the space of such $X_i$s are one point for
$n<N$ and $\Cplx^2/\Intg_N$ for $n=N$ \cite{DM,DDG}. Because the moduli
spaces of noncommutative solitons are determined by the $X_i$
\cite{KMT}, only when the solitons of all $N$ types coincide they can 
move from the origin.

\subsection{From D4-\={D}4 to D2}
\label{D4-D2}

There are many ways to obtain system with an $\alpha$ type D4-brane and
a $\beta$ (assume $\beta\geq\alpha$) type \=D4-brane using the solution
generating technique. Here we give some examples.

\subsubsection{Simple $\alpha$-$\bar{\beta}$ system}

The most simple $\alpha-\bar{\beta}$ systems can be obtained by solution
generating technique from $(\beta-\gamma)$-$\overline{(\beta-\gamma)}$
system, where $\gamma$ satisfying $0\leq\gamma\leq\beta-\alpha$ is an
integer. If we act with the shift operator $S_1^{\beta-\alpha-\gamma}$
on the  D-brane and $S_2^{\gamma}$ on the \={D}-brane, we obtain
$\alpha$-$\bar{\beta}$ systems for which
\begin{align}
 \begin{split}
  T&=\Sb_1^{\beta-\alpha-\gamma}\otimes S_2^{\gamma} \in\Mod{\alpha}{\beta},\\
  C_1^+&=S_1^{\beta-\alpha-\gamma}\crea_1\Sb_1^{\beta-\alpha-\gamma},
  \Cb_1^+=S_1^{\beta-\alpha-\gamma}\anih_1\Sb_1^{\beta-\alpha-\gamma},\\
  C_2^-&=S_2^\gamma\crea_2\Sb_2^\gamma,\Cb_2^-=S_1^\gamma\anih_2\Sb_2^\gamma,\\
  \Tb T&=(I-P_{\beta-\alpha-\gamma})\otimes I, \\
  T \Tb&= I\otimes (I-P_{\gamma}).
 \end{split}
\end{align}
Following the identification rules given in \S\ref{frac}, we
get $\alpha+1,\ldots,\beta-\gamma$ type D2-branes that
extend in the $z_2$ plane and $\beta-\gamma+1,\cdots,\beta$ type
D2-branes extend in the $z_1$ plane. In any case,
$\beta-\alpha$ D2-branes are created. The calculation in the
$\gamma=0$ case has been carried out in the context of BSFT
\cite{Tak1} and we have the same
number of D2-branes, while the types of these D2-branes are the same.

The gauge field on the type $\alpha$ D4-brane is given by
\begin{align}
 C^+_1&=S_1^{\beta-\alpha-\gamma} \crea_1 \Sb_1^{\beta-\alpha-\gamma},\quad
 \Cb^+_1=S_1^{\beta-\alpha-\gamma} \anih_1 \Sb_1^{\beta-\alpha-\gamma}.
\end{align}
The curvature on the $\alpha$ brane is
\begin{equation}
 -i\theta^{-1}([C_1,\Cb_1]+I)=-i\theta^{-1}P_{\beta-\alpha-\gamma}\otimes I.
\end{equation}
Integrating over the $z_1$ plane, we have the winding number $(\beta -\alpha
-\gamma)/N$. Similarly we have winding number $\gamma/N$ around the $z_2$ plane
on the $\bar{\beta}$ brane. This means that the $\alpha$ brane has
D2-brane charge$(\beta-\alpha-\gamma)$  and $\bar{\beta}$
brane has D2-brane charge $\gamma$.

We can verify the above result by calculating the tension. The energy of
the system can be calculated by the action for the D4-brane
(\ref{D4-D4bar}). The kinetic terms for the gauge fields vanish because the
coefficient $f(T,\Tb)$ vanishes. The kinetic terms for tachyon field
also vanish, because these terms are zero before the solution
generating. Thus it is sufficient to examinine the potential term;
\begin{align}
 \Lag= T_{4}\left(\frac{2\pi\ap}{\theta}\right)^2 \frac{(2\pi\theta)^2}{N}
 \Tr\big(V(1-T\Tb)+V(1-\Tb T)\big)
 =T_2\frac{2\pi\ap}{\theta}V_2(\beta-\alpha),
\end{align}
where $V_2$ is volume of $z_1$ or $z_2$ plane. Thus the tension is
$T_2 (2 \pi \ap / \theta)(\beta-\alpha)$, which is the correct value for
the $\beta-\alpha$ D2-branes.

\subsubsection{Additional D2-\=D2 pairs creation}

In the $\alpha$-$\bar{\beta}$ system, there is another type of the tachyon
configuration that enables us to obtain the D2-\=D2 system.
Consider the case that we act with the shift operator $S_2^\gamma$ on the
D-brane and with $S_2^{\beta-\alpha+\gamma}$ on the \={D}-brane for the
$(\alpha-\gamma)$-$\overline{(\alpha-\gamma)}$ system, where $\gamma$ is
a positive integer. We then obtain $\alpha$-$\bar{\beta}$ system ,where
\begin{align}
 \begin{split}
  T&=I\otimes S_2^{\beta-\alpha+\gamma}\Sb_2^{\gamma}\in\Mod{\alpha}{\beta},\\
  C_2^+&=S_2^{\gamma}\crea_2\Sb_2^\gamma,
  \Cb_2^+=S_2^\gamma\anih_2\Sb_2^\gamma,\\
  C_2^-&=S_2^-{\beta-\alpha+\gamma}\crea_2\Sb_2^{\beta-\alpha+\gamma},
  \Cb_2=S_2^{\beta-\alpha+\gamma}\anih_2\Sb_2^{\beta-\alpha+\gamma},\\
  T \Tb &=I\otimes(I-P_{\beta-\alpha+\gamma}), \\
  \Tb T &=I\otimes(I-P_{\gamma}).
 \end{split}
\end{align}
We have $\alpha-\gamma+1,\ldots,\beta$ type \={D}2-branes and
$\alpha-\gamma+1,\ldots,\alpha$ type D2-branes. Since
D2-branes and \=D2-branes of same type annihilate, this system
will be pair annihilate, yielding $\alpha+1,\ldots,\beta$ type
\={D}2-branes. We note that this is consistent with the claim that
the topological charge is determined by the asymptotic behavior of
the tachyon field. Upper-left part of the matrix represents the behavior
near the origin and the (infinitely) lower-right part represents the
asymptotic behavior far from the origin. As is easily seen from the
explicit form of $T$, asymptotic behavior does not depend on the value
of $\gamma$ but, rather is determined by the value of $\beta-\alpha$:
\begin{align}
 \begin{split}
  T=S_2^{\gamma}\Sb_2^{\beta-\alpha+\gamma}=I\otimes
  \bordermatrix{
       & 0    &      &\beta-\alpha+\gamma &        &      &\\
 0     & 0    &\cdots& 0                  & \cdots &\cdots&\cdots\\
       &\vdots&\ddots&\ddots              & \ddots &\ddots&\ddots\\
 \gamma&\vdots&\ddots& 1                  & 0      & 0    &\ddots\\
       &\vdots&\ddots& 0                  & 1      & 0    & 0    \\
       &\vdots&\ddots& 0                  & 0      & 1    & 0    \\
       &\vdots&\ddots& \ddots             & 0      & 0    &\ddots\\}.
 \end{split}
\end{align}
Thus the total R-R charge does not depend on $\gamma$.

\subsection{From D2-\={D}2 to D0}

This case is almost the same as that in the previous subsection, but
here we must work only in the $z_1$ plane or the $z_2$ plane. Thus we
must restrict the operators which appear in the calculation to the
form $I\otimes M$ or $M\otimes I$.  From the $\alpha$ type D2-brane
and $\beta$ type \={D}2-brane we get $\alpha+1 ,\ldots, \beta$ type
D0-branes.

\subsection{From D4-\={D}4 to D0}

In the commutative case, to see the decay mode by which the  D4-\=D4 system decays
into D0 system, we must consider two pairs of D4-\=D4. Therefore we will
modify the operator representation of the start point of the solution
generating technique by tensoring the CP-factor,
$\begin{pmatrix} 1 & 0 \\ 0 & 1
 \end{pmatrix}$. 
In this way, we can calculate two general pairs of D4-\=D4 systems.

\subsubsection{ABS-like construction}
\label{ABS-like}

To consider the codimension four case, we will begin with the
relatively simple configuration of the $(\alpha,\beta+\delta)$-$
(\overline{\alpha+\delta},\bar{\beta})$ system. The reason for
considering this configuration is that this configuration allows the
ABS(Atiyah-Bott-Shapiro)-like construction \cite{Wit1,HM,Tak1} where
only D0-branes are created as we will see below. The tachyon field is
written
\begin{align}
 \begin{split}
  T&\propto\begin{pmatrix}
	    \bar{T_2} & T_1 \\
	    \bar{T_1} & -T_2
	   \end{pmatrix},\\
  T_1&=S_1^{-\gamma_1}\otimes
  S_2^{-\gamma_1+\alpha-\beta}\in\Mod{\alpha}{\beta},\\
  T_2&=S_1^{-\gamma_2}\otimes S_2^{-\gamma_2-\delta}\in\Mod{0}{\delta},
 \end{split}
\end{align}
where $\gamma_1$and $\gamma_2$ are any integer and negative power of the
$S_i$ means $S_i^{-\alpha}=\Sb_i^{\alpha}$. The proportionality here means that $T$
must be normalized by the square root of $\Tb T$ from the right. As the
solutions of the equations $\Tb T\Tb=\Tb$ and $T\Tb T=T$, we found solutions
$(\gamma_1 , \gamma_2) = ( \alpha - \beta , 0),( 0, -\delta )$. In this
case, we have
\begin{align}
 \begin{split}
  \Tb T =\begin{pmatrix}
	   I\otimes I & 0 \\
	   0 & I \otimes I
	  \end{pmatrix},\quad
  T \Tb =\begin{pmatrix}
	   I\otimes I & 0 \\
	   0 & I\otimes I-P_{\beta-\alpha}\otimes P_\delta
	  \end{pmatrix},
 \end{split}
\end{align}
for $(\gamma_1 , \gamma_2)=(\alpha - \beta ,0)$, and
\begin{align}
 \begin{split}
  \Tb T =\begin{pmatrix}
	   I\otimes I & 0 \\
	   0 & I \otimes I
	  \end{pmatrix},\quad
  T \Tb =\begin{pmatrix}
	   I\otimes I- P_\delta\otimes P_{\beta-\alpha} & 0 \\
	   0 & I\otimes I
	  \end{pmatrix},
 \end{split}
\end{align}
for $(\gamma_1 , \gamma_2)=(\alpha - \beta ,0)$. These operators
are apparently projection operators.

The solution $(\gamma_1,\gamma_2)=(\alpha-\beta,0)$ gives a
$\delta(\beta-\alpha)$ fractional \={D}0-brane on the
$\overline{\alpha+\delta}$ brane and the types of the \={D}0-branes are
\begin{equation}
 \begin{matrix}
  \alpha+\delta+1&,\ldots,&\beta+\delta\\
  \vdots&\ddots&\vdots\\
  \alpha+2&,\ldots,&\beta+1
 \end{matrix}.\label{D4toD0}
\end{equation}
Because only \={D}0-branes exist, $1/2$ supersymmetry is preserved after
the tachyon condensations. The solution
$(\gamma_1,\gamma_2)=(0,-\delta)$ gives the same result. This result
gives same number and types of D0 branes as the result of Ref 26).
The solution $(\gamma_1,\gamma_2)=(0,-\delta)$ gives
$\delta(\beta-\alpha)$ fractional \={D}0-branes on the $\bar\beta$ brane
and the type of the \={D}0-brane is the same as that in the case
$(\gamma_1,\gamma_2)=(\alpha-\beta,0)$

The gauge field on the \=D4-brane in this case is
\begin{align}
 \begin{split}
  C^-_{1,2}&\propto T\begin{pmatrix}
	      \crea_{1,2} & 0\\
	      0 & \crea_{1,2}     
	     \end{pmatrix}\Tb,\\
  \Cb^-_{1,2}&\propto T\begin{pmatrix}
		 \anih_{1,2} & 0\\
		 0 & \anih_{1,2}     
		\end{pmatrix}\Tb.
 \end{split}
\end{align}
The gauge field on the D4-brane is trivial. The curvature on the \=D4-branes
is
\begin{equation}
 -i\theta^{-1}\big([C_i^- ,\Cb_i^-]+I\big)=\left\{
					    \begin{matrix}
		 -i\theta^{-1}\begin{pmatrix}
			       P_{\beta-\alpha}\otimes P_\delta & 0 \\
			       0 & 0
			      \end{pmatrix}
		\quad\mbox{for} (\gamma_1,\gamma_2)=(\alpha-\beta,0),\\
		-i\theta^{-1}\begin{pmatrix}
			      0 & 0 \\
			      0 & P_\delta\otimes P_{\beta-\alpha}
			     \end{pmatrix}
		\quad\mbox{for} (\gamma_1,\gamma_2)=(0,-\delta).
					    \end{matrix}\right..
\end{equation}
The winding number is $\delta(\beta-\alpha)/N$. This means \={D}4-branes
have a $\delta(\beta-\alpha)$ D0-brane charge.

\subsubsection{Decay of two general pairs of D4-\=D4}

We can extend the above calculation to the
$(\alpha,\beta+\delta')$-$(\overline{\alpha+\delta},\bar{\beta})$
system. This is the general two pairs of D-\={D} system. Without loss of
generality, we can $\delta'\geq\delta$. We also choose the tachyon field
as
\begin{align}
 \begin{split}
  T&\sim\begin{pmatrix}
      \Tb_2& T'_1 \\
      \Tb_1& -T_2'
     \end{pmatrix},\\
  T_1&=S^{\beta-\alpha}\otimes I, \\
  T'_1&=S^{\beta-\alpha}\otimes\Sb^{\delta'-\delta}, \\
  T_2&=I\otimes\Sb^\delta, \\
  T'_2&=I\otimes\Sb^{\delta'}.
 \end{split}
\end{align}
This system decays into anti-D0-branes of type described by
(\ref{D4toD0}) and D2-branes extending in the $z_1$ plane of type
$\delta+1,\ldots,\delta'$. This is consistent with the result of the charge calculation
using a boundary state. Furthermore, the existence of the D2-brane breaks
$1/2$ supersymmetry.

The gauge field is the same as that in the case of \S\ref{ABS-like} for
anti-D4-branes and
\begin{equation}
 C= \begin{pmatrix} 
     \crea & 0 \\ 0 & S_2^{\delta'-\delta}\crea\Sb_2^{\delta'-\delta}
    \end{pmatrix},\quad
    \Cb = \begin{pmatrix} 
	   \anih & 0 \\ 0 & S_2^{\delta'-\delta}\anih\Sb_2^{\delta'-\delta}
	  \end{pmatrix}
\end{equation}
for D4-branes. The winding number of the gauge field around the $z_2$ plane on
D4-branes is $(\delta' - \delta)/N$. This means that a $\delta'-\delta$
D2-brane charge exists.

\section{Conclusion and discussion}
\label{conclusion}

In this paper, we have summarized the construction of the
noncommutative field theory on the noncommutative orbifold
$\Cplx^2/\Intg_N$ following the prescription of the Ref. \cite{MM} and
proposed identification rules of the noncommutative solitons as
fractional branes. Those rules determine the correspondence between
the types of the noncommutative solitons and these of the fractional
branes. We showed that the noncommutative solitons on the orbifolds
have the same nature -- such as fractional tension and pinning -- as
those on (fractional) D-brane.

The solution generating technique was extended to give more solutions in
the orbifold theory. Combined with our identification rules, this
method provide powerful tools to analyze nonperturbative aspects of
string theory on orbifolds.

As examples, by calculating the various decay modes of D-\={D}
systems on noncommutative orbifolds, we obtained results consistent with the
result calculated using BSFT. Furthermore, using the extended
solution generating technique, we were able to calculate the result in the case of more
general decay modes of D-\={D} systems on noncommutative
orbifolds. In particular, we derived the decay mode of two D4-\={D}4 pairs,
which decay into D0-branes and D2-branes. The extension of this
calculation to any pairs of D4-\={D}4 is straightforward.

As further problems, we must confirm the identification given in
\S\ref{frac} more rigorously. We determined some signs in the formula
that determines the type of D0-brane and D2-brane so that the
calculation in \S\ref{condensation} give results consistent with the
result obtained using BSFT. But there should be the reason for these
signs.

Furthermore we need a more detailed derivation showing why the redefinition of the
algebra in \S\ref{OrbAlg} has some physical meaning. The
properties of the redefined algebra have important roles in the
correspondence between the noncommutative solitons and D-branes. As we
stated in \S\ref{OrbAlg}, we believe that this is because string
theory ``sees'' the covering space of the orbifold by including the twisted
sectors.

Furthermore, although we concentrated on $\Cplx^2/\Intg_N$, we can
calculate non-abelian orbifolds such as $\Cplx^2/\Gamma$, where $\Gamma$
is $D_n$ or $E_{6,7,8}$ or higher dimensional orbifold \cite{Tak1}.
These calculations will provide a good check of our identifications given
in \S\ref{frac}.

\subsection*{Acknowledgments}

I am very grateful to Y. Matsuo and T. Takayanagi for many advices, very
useful discussion and encouragement. I also thank S. Terashima, K. Sakai,
K. Ohmori, Y. Imamura, and Y. Konishi for useful discussions, and
M. Hamanaka, and M. Nozaki for reading drafts and encouragement. This work
is supported by JSPS Research Fellowships for Young Scientists.

\end{document}